\documentclass[10pt, conference, letterpaper]{IEEEtran}

\usepackage{longtable}
\usepackage{epsfig,endnotes}
\usepackage{booktabs}
\usepackage{amsmath}
\usepackage{pifont}
\usepackage{multirow}
\usepackage{url}
\usepackage{algorithm}
\usepackage[noend]{algpseudocode}

\newcommand\ignore[1]{}
\newcommand\ignoreInfocom[1]{}

\begin{document}

\title{IoT or NoT: Identifying IoT Devices in a Short Time Scale}

\author{
 Anat Bremler-Barr, Haim Levy, Zohar Yakhini \\
  \IEEEauthorblockA{Computer Science Department, Interdisciplinary Center, Herzliya, Israel}

}

\ignore{ \author{
{\rm Anat Bremler-Barr} \\
IDC
\and
{\rm  Haim Levy} \\
IDC
\and
{\rm  Zohar Yakhini}\\
IDC}
}

\newcommand{\hl}[1]{{\sffamily\bfseries \mbox{***HL:\ } #1 \mbox{****}}}
\newcommand{\ab}[1]{{\sffamily\bfseries \mbox{***ABB:\ } #1 \mbox{****}}}

\maketitle
\subsection*{Abstract}

In  recent  years the number of IoT devices in home networks has increased dramatically. Whenever a new device connects to the network, it must be quickly managed and secured using the relevant security mechanism or QoS policy. Thus a key challenge is to distinguish between IoT and NoT devices in a matter of minutes. Unfortunately, there is no  clear indication of whether a device in a network is an IoT.  In this paper, we propose different classifiers that identify a device as IoT or non-IoT, in a short time scale, and with high accuracy.  

 Our classifiers were constructed using machine learning techniques on a seen (training) dataset and were tested on an unseen (test) dataset. 
 They successfully classified devices that were not in the seen dataset with accuracy above $95\%$. 
The first  classifier is a logistic regression classifier based on traffic features. The second classifier is based on features we retrieve from DHCP packets. Finally, we present a unified classifier that leverages the advantages of the other two classifiers. 
We focus on the home-network environment, but our classifiers are also applicable to enterprise networks.

\section{Introduction}
\label{introduction}

The number of IoT devices in home network has increased dramatically in recent years.  IoT devices are much more vulnerable to attacks than general-purpose endpoint computers and will be insecure in the foreseeable future. In most cases, the IoT device is strong enough to host an attacking zombie but too weak to protect itself from malicious code. Thus, it clearly  poses new security challenges.  Attacks on IoTs have severe implications in both the cyber and physical domains \cite{DDoS-FI,Yu2015,Wei2018,kolias17}.  There are many proposed security and management solutions, with the common practice being a network-based solution that is geared to protecting IoT devices and resides at the home router \cite{MCAFEE, ASUS} or in an additional security device designed to protect the home network \cite{rattrap, cujo, Fingbox, bitdefender}.  The security solution cannot reside in the IoT device itself, due to the low CPU power and memory of IoT. 
  
  Whenever a new device connects to the network, it must be managed and secured as quickly as possible using the relevant security policy.  
A key challenge is thus to quickly distinguish between IoT (smart camera, bulbs, speakers and so on) and non-IoT devices( general purpose computers,  mobile phones, desktops, tablets and laptops and so on), referred to in our paper as NoT devices.  We assume that the classification is done in a device that observes the LAN traffic. Our paper focuses on IoTs that are actual physical entities connected to the internet. Classification of borderline devices such as smartTVs (that  can be used to surf the Internet and to run different applications) is part of our future work.

In this paper, we propose three classifiers for identifying devices as IoT or NoT (see a summary of results in Table \ref{tab:final_results}). We use a  machine learning methodology, where we trained  classifiers on the \emph{seen} dataset of labeled (IoT or NoT)  devices, and then analyze the accuracy of our classifiers on an \emph{unseen} dataset of devices. The unseen dataset includes IoT device types that \emph{ were not in the seen dataset}. This  challenging requirement is due to the huge variety of IoT devices,   types and  vendors. Moreover, the IoT market is very dynamic, with new vendors and new devices appearing constantly. We therefore seek  a general classifier: one that  identifies the inherent characteristics of IoT vs NoT devices.   
 
 The desired classification approach should have the following properties:
\begin{itemize}
    \item{Universality}  - the classifier should be generic and work for all IoT device types, including those with encrypted traffic. 
    It should also work on device types it has not yet seen. 
    
    \item{Low classification latency} - We consider 1 to 20 minute latency.
    
    \item {Accuracy} - the classifier should be accurate.  We use  F1-score, recall and precision as measures for accuracy.
   
    \item{Efficient} - the classifier should also be efficient, with low CPU processing and memory requirements, since it needs to process on-line traffic. 
    
     \item{Passiveness} - the classifier should passively process the traffic and may not use active probing of devices. Active techniques are usually tailored for specific IoT devices and/or require special permission from the owner. Moreover, in some cases, active probing might unintentionally activate the devices.
\end{itemize}

\begin {table*}[ht]
\begin{center}
\resizebox{\linewidth}{!}
{
\begin{tabular}{|l||l|l|l|l|l| }
\hline
Classifier	& Classification Latency &	 F1-Score & Recall &  Precision & Efficiency\\
\hline
\hline

Classifier I: Traffic Features & 1 min & 91.54\% & 97.38\% & 87.02\%  & Counters\\
 & 5 min & 96.59\% & 98.72\% & 94.54\%  & Counters\\
 & 10 min & 98.12\% & 98.91\% & 97.34\% & Counters\\
 & 20 min & 98.07\% & 98.71\% & 97.43\%   & Counters\\
\hline
Classifier II: DHCP & Long & 95.73\% & 93.75\% & 97.82\% &  DPI required\\
\hline
Classifier III: Unified & 20 min & 99.04\% & 98.11\% &100\% & DPI required \\
\hline
\end{tabular}}

\vskip 6pt
\caption{Experimental results of  the presented classifiers over the unseen dataset \vspace{-0.5cm}}  \label{tab:final_results} 
\end{center}
\end {table*}
\smallskip

The first classifier, presented in Section \ref{sec:TrafficClassifier}, is a logistic regression classifier using {\bf traffic features}. The resulting classifier is efficient and requires light processing on a few features  of the traffic (up to 5). We found the most informative features to be the  TCP window size and the number of unique DNS queries. We show that these features were chosen by our ML algorithm because of the very limited number of endpoints with which IoT devices communicate, as well as their small TCP buffer size. 

The second classifier, presented in Section \ref{sec:DHCP}, is based on a  decision tree on {\bf DHCP} information. The DHCP protocol is very common in home networks, and the resulting decision tree is simple (with a height smaller than 5). The downside of the algorithm is that not all the networks or devices use DHCP. Moreover, it might have a long classification latency, since DHCP appears relatively seldom in the traffic. 

We then present in Section  \ref{sec:unif_res} a  {\bf unified classifier} 
that leverages the advantages of the other two above classifiers and achieves F-score, precision and recall above $98$. The  unified classifier can be found in our github repository \cite{GITHUB}. 

The   closest  work to ours is the recent published work, DeviceMien \cite{Ortiz2019}, where the authors also note the blind spot in the literature in categorizing as IoT previously unseen IoT devices. However we achieve better accuracy, and we also provide  a clear intuition behind the features and signatures selected by our classifiers (see Section \ref{sec:related}). Thus, our work sheds light on the unique network characteristics of IoT devices. 

\section{Related work}
\label{sec:related}

A technique that uses user-agent field information was suggested in \cite{Meidan2017} to classify IoT devices. A user-agent value is sent during HTTP requests, and it  contains a short description of the properties of the requesting device. For NoT devices, this parameter is usually of greater length, since  it describes properties  relevant only for NoT devices, such as screen size, OS language and browser. 
However, we discovered that the technique does not meet our requirements. The user-agent parameter cannot be observed in encrypted traffic. In our dataset, only 69.5\% of devices transmit this parameter as a plain text, and a similar result was shown in \cite{Lastovicka2018}. Moreover classification latency is  high - the chance to find an user-agent value sent by an IoT device in a slot of 20 minutes is about 25\%. We note that unlike the DHCP client information, the user- agent is information that was sent to the endpoints, and hence it is not stored at the home-router and thus cannot be actively retrieved. 

MAC OUI can be helpful in identifying the manufacturer of a  device. But is of very limited use as a unique identifier of IoT due to manufacturers that supply both IoT and non-IoT device types (such as  Samsung's smartcam and smartphones). The authors of \cite{Martin2016} tried to associate MAC ranges of manufacturers with models,  but  their technique was often ineffective   due to lack of regularization in this field.

Other works address related areas, such as \emph{device} fingerprinting \cite{Sivanathan2017, Guo2016, Meidan2017, Msadek2019}, but  IoT devices   that were not seen before cannot be identified by these techniques. 

\ignoreInfocom{
However, the MAC OUI is of very limited use as a unique identifier for our cause due to manufacturers that supply both IoT and non-IoT device types (such as  Samsung's smartcam and smartphones) and those that supply generic multiple-purpose chipsets which are being spread to numerous purposes. In about 20\% of the IoT devices we examined, the OUI corresponds to the NIC manufacturer (e.g., AmpakTec for Pixstar Photo Frame, or Shenzhen Reecam for Insteon Camera).}

The authors of \cite{Feng2018} propose a proactive request to IoT devices in order to classify the exact IoT vendor/model device. However, this work does not meet our passiveness  requirement, since sending packets to IoT requires non-trivial permissions.

OS fingerprinting is addressed in \cite{Kollmann2007, Kollmann, Lastovicka2018}; however, the device type cannot be easily identified from the resulting OS information.  The Satori project \cite{Kollmann2007, Kollmann} inspired our use of some of the features we tested in this work, mostly data from the IP-TCP layers.

The   closest  work to ours is DeviceMien \cite{Ortiz2019}, where the authors also note the blind spot in the literature in categorizing as IoT previously unseen IoT devices. Their approach, which  uses auto-encoder,  a deep-learning technique, cannot provide any intuition regarding the received classifier,  in contrast to our approach.  We also achieve better accuracy:   F1-score of above $95\%$ on ours dataset, as opposed to $76\%$ by the authors of \cite{Ortiz2019} on theirs.
 Since they did not publish the dataset or their resulting classifier,  we cannot compare our techniques on the same dataset. However, they did provide a list of the devices in the dataset, which are very similar to our. We note that they also checked a few borderline IoT devices, such as SmartTVs. We predict, based on the list of devices, that we would achieved an F1-score of $92\%$ on their dataset, assuming misclassification in the borderline IoT devices.  We suspect that our superior results due to the ability of machine learning techniques such as the one we used to achieve good results on small datasets. Deep learning techniques, on the other hand, require huge datasets, which are difficult to obtain due to the need to label the devices. Another advantage of our classifiers is that they latency is time-bound. Finally, our implementation is more efficient since they require only a few memory references.

We note that a recent initiative calls for IoT device vendors to provide a \emph{Manufacturer Usage Description (MUD)} for their IoT products~\cite{ietf-opsawg-mud-25}, which as a by-product identifies the device as an IoT. Only a very few IoTs currently  provide MUD files. It is moreover questionable whether the majority of the vendors would comply with MUD, since the vendor apathy is one of the root causes of the IoT security problems.

\section{Methodology}
\label{sec:methodology}

Our dataset is composed of captured network traffic data (pcap files), recorded at the router or at an access point, of labeled devices from various sources: \cite{Sivanathan2017}, \cite{cicids2017} , \cite{cicids2012}, \cite{lander}, \cite{Alrawi2019} and pcaps collected from our IoT lab. The IoT devices in our dataset are unique by type, model and/or OS version.
Overall we had about 46GB of data. Recording time varied greatly, with some devices that were recorded for weeks and some for hours. Overall, our dataset contained 121 devices: 77 IoT and 44 NoT devices.

At first, we  arbitrarily split the dataset into two groups: \emph{seen} and \emph{unseen}. Later on, we gained access to more devices, and we added them to the unseen dataset. Our seen group contained 45 devices, 24 IoT and 21 NoT (see Table \ref{Tbl:seen} in Appendix). Our unseen group contained 76 devices, 53 IoT devices and 23 NoT (see Table \ref{tbl:unseen} in Appendix). 

As the names indicate, we performed the learning on the seen group and tested our classifiers on the unseen group. We want to emphasis that the samples of the unseen group were not available to us during the learning phase.

In order to test classification performance in various time periods, we divided our dataset into time slots of 1,  5, 10 and 20 minutes. Working with a model of slots, our classifiers  process information of a time slot (in the training phase and testing phase). Slots with a small number of packets were also considered, since we  observed that they might contain sufficient information for classification.  Some devices were characterized by their very rare network usage, such as the Nest Smoke Alarm, which sends only a few packets every 23 hours. Thus, our classifier can classify a device as soon as it sends data.

Our goal is to classify a new device in a network as being  IoT or NoT. This goal of dichotomy classification is a choice we made. We could also have used a scoring mechanism, estimating the probability of being in one of the classes, or classification to three categories: IoT or NoT or Undecidable. The dichotomy classification fits well with our need to always decide how to protect and manage  the device, as a general purpose computer or as an IoT device.

In our model, an IoT device is considered 'Positive', and a NoT device is considered 'Negative'. We thus defined the following performance metrics:  True Positive (TP) -  correct classification of an IoT device; False Positive (FP) - misclassification of a non-IoT (NoT) device as IoT  True Negative (TN) - correct classification of a non-IoT (NoT) device;  False Negative (FN) - misclassification of an IoT device as non-IoT (NoT). 

We measure the accuracy using recall, precision and F1-Score: {\bf Recall} is the probability of an actual IoT to be successfully classified as such, i.e., $\frac{TP}{TP+FN}$. {\bf Precision}  is the probability that an IoT-classified device is truly an IoT, i.e., $\frac{TP}{TP+FP}$. {\bf F1-score} is a unified performance index defined as  $2 \cdot \frac{recall \cdot precision }{recall+precision}$.

\ignore{
\begin{itemize}
    \item True Positive (TP) -  correct classification of an IoT device 
    \item False Positive (FP) - misclassification of an non-IoT (NoT) device as IoT.
    \item True Negative (TN) - correct classification of a non-IoT (NoT) device. 
    \item False Negative (FN) - misclassification of an IoT device as non-IoT (NoT). 
\end{itemize}

We measure the accuracy using recall, precision and F1-Score.

\begin{description}
    \item[Recall] is the probability of an actual IoT to be successfully classified as such, i.e., $\equation{ \frac{TP}{TP+FN}}$.
    \item[Precision] is the probability that an IoT-classified device is truly an IoT, I.e., $\equation{ \frac{TP}{TP+FP}}$.
    \item[F1-score] is a unified performance index defined as  $\equation{ 2 \cdot \frac{recall \cdot precision }{recall+precision}}$
\end{description}
}

\ignoreInfocom{
\section{Appendix}

{\tt

\begin{table}
\scriptsize
\centering
\small
\begin{tabular}{|l||l| }
\hline
 Category & Device Name    \\  
\hline
\hline
\multirow{21}{*}{NoT}  & Samsung Galaxy Tab    \\
    &  Android Phone      \\ 
    &  Windows Laptop     \\ 
    &  MacBook    \\ 
     &  Android Phone     \\ 
     &  IPhone    \\ 
     &  MacBook/Iphone     \\ 
     &  Lg Smartphone    \\ 
     &  Nexus 5x Smartphone    \\ 
     &  Apple Macbook    \\ 
     &  Xiaomi Smartphone     \\ 
     &  Laptop win10    \\ 
    &  Macbook    \\ 
    &  Macbook     \\ 
    &  Samsung s7    \\ 
    &  Xiaomi mi5    \\ 
    &  Galaxy-A7-2017 Smartphone    \\ 
     &  Lenovo Win10 laptop    \\ 
     &  LG G3 Smartphone    \\ 
     &  Asus ZenPad Tablet    \\ 
     &  Dell Win7 laptop     \\
     \hline
  \multirow{24}{*}{IoT}  &  Smart Things  \\ 
     &  Netatmo Welcome    \\ 
     &  TP-Link Day Night camera   \\ 
      &  Samsung SmartCam    \\ 
    &  Dropcam    \\ 
    &  Insteon Camera, wifi   \\ 
      &  Insteon Camera, wired   \\ 
      &  Withings Smart Monitor   \\ 
     &  Belkin Wemo switch    \\ 
      &  TP-Link Smart plug    \\ 
     &  iHome    \\ 
      &  Belkin wemo motion sensor    \\ 
      &  NEST Protect smoke alarm     \\ 
     &  Netatmo weather station     \\ 
     &  Withings Smart scale     \\ 
     &  Blipcare Blood Pressure meter     \\ 
    &  Withings Aura smart sensor     \\ 
   &  LiFX Smart Bulb     \\ 
      &  Triby Speaker Smart Speaker  \\ 
    &  PIX-STAR Photo-frame    \\ 
     &  HP Printer     \\ 
     &  Nest Dropcam    \\ 
    &  Chromecast Streamer  \\ 
     &  Yeelink Smart Light Bulb    \\
 \hline
\end{tabular}
\caption{ Seen Dataset \label{Tbl:seen}}
\end{table}
}

{\tt 
\begin{table}
\scriptsize
\centering
\small
\begin{tabular}{|l||l|}
\hline
  Category & Device Name    \\  
\hline
\hline
 \multirow{23}{*}{NoT} &   Samsung s5 Smartphone  \\ 
      &  Android samsung      \\ 
      &  Apple iPhone     \\ 
      &  Windows Laptop     \\ 
      &  Ubuntu PC    \\ 
      &  Apple ipad    \\
      &  xiaomi A2     \\
      &  Dell Laptop win 10    \\
      &  Mac laptop    \\
      &  Xiaomi smartphone     \\   
      &  VM Win 8.1 64B     \\ 
      &  VM Win 7 Pro 64B     \\ 
      &  VM Ubuntu 16.4 64B     \\ 
      &  VM Win 10 pro 32B    \\ 
      &  VM Ubuntu 16.4 32B     \\ 
      &  VM Ubuntu 14.4 64B     \\ 
      &  VM Ubuntu 14.4 32B    \\ 
      &  Macbook   \\  
      &  VM Testbed09 (windows)   \\ 
      &  VM Testbed13 (windows)     \\ \ 
    & iPad   \\
    & Iphone   \\
   & Android Tablet   \\
\hline
\multirow{53}{*}{IoT}
& 2X Amazone Echo \\
& Apple HomePod \\
& August Doorbell Cam \\
& Belkin Netcam \\
& Belkin WeMo Link \\
& Bezeq smarthome \\
& Bose SoundTouch 10 \\
& Canary \\
& Caseta Wireless Hub \\
& Chamberlain myQ Garage Opener \\
& Chinese Webcam \\
& D-Link DCS-5009L Camera \\
& Foscam \\
& Google Home \\
& 3X Google Home Mini \\
& Google OnHub \\
& Harmon Kardon Invoke \\
& Insteon Hub \\
& iRobot \\
& Koogeek Lightbulb \\
& lifiLab \\
& Logitech Harmony Hub \\
& Logitech Logi Circle \\
& MiCasaVerde VeraLite \\
& 2X Motorola Hubble \\
& NestCam \\
& NestDetector \\
& Nest Camera \\
& Nest Cam IQ \\
& Nest Guard \\
& Netgear Arlo Camera \\
& Philips HUE Hub \\
& Piper NV \\
& Provision ISR \\
& RENPHO Humidifier \\
& Ring Doorbell \\
& Roku 4 \\
& Roomba \\
& samsung smart home camera \\
& Securifi Almond \\
& smartHub \\
& Sonos \\
& TP-Link Smart WiFi LED Bulb \\
& 2X TP-Link WiFi Plug \\
& WeMo Crockpot \\
& Wink 2 Hub \\
& Withings Home \\
& Wyze Cam \\

\hline
\end{tabular}
\caption{ Unseen Dataset  \label{tbl:unseen}}
\end{table}
}

\ignore{
{\tt
\begin{table*}
\scriptsize
\small
\begin{tabular}{|l||l|l|||l|l}
\hline
Category & Device Name (entered label)& Database  & Device Name (entered label)& Database \\
\hline
\hline
\multicolumn{3}{c}{Seen} & \multicolumn{2}{c}{Unseen} \\
\hline
\multirow{21}{*}{NoT}  & Samsung Galaxy Tab         & \multirow{7}{*}{\cite{Sivanathan2017}} & Samsung s5 Smartphone &  \multirow{10}{*}{Our volunteers} \\
                       &  Windows Laptop            &                                        & Apple iPhone          &  \\
                       &  Android Phone             &                                        & Android samsung       &  \\
                       &  MacBook                   &                                        & Windows Laptop        &  \\
                       &  Android Phone             &                                        & Ubuntu PC             &  \\
                       &  IPhone                    &                                        & Apple ipad            &  \\
                       &  MacBook/Iphone            &                                        & xiaomi A2             &  \\
\cline{2-3}
                       &  Lg Smartphone             &  \multirow{14}{*}{Our volunteers}      & Dell Laptop win 10    &  \\
                       &  Nexus 5x Smartphone       &                                        & Mac laptop            &  \\
                       &  Apple Macbook             &                                        & Xiaomi smartphone     &  \\
\cline{4-5}
                       &  Xiaomi Smartphone         &                                        & VM Win 8.1 64B        &  \multirow{8}{*}{\cite{cicids2017}} \\
                       &  Laptop win10              &                                        & VM Win 7 Pro 64B      &  \\
                       &  Macbook                   &                                        & VM Ubuntu 16.4 64B    &  \\
                       &  Macbook                   &                                        & VM Win 10 pro 32B     &  \\
                       &  Samsung s7                &                                        & VM Ubuntu 16.4 32B    &  \\
                       &  Xiaomi mi5                &                                        & VM Ubuntu 14.4 64B    &  \\
                       &  Galaxy-A7-2017 Smartphone &                                        & VM Ubuntu 14.4 32B    &  \\
                       &  Lenovo Win10 laptop       &                                        & Macbook               &  \\
\cline{4-5}
                       &  LG G3 Smartphone          &                                        & VM Testbed09 (windows) &  \multirow{2}{*}{\cite{cicids2012}} \\
                       &  Asus ZenPad Tablet        &                                        & VM Testbed13 (windows) &  \\
\cline{4-5}
                       &  Dell Win7 laptop          &                                        &                        &  \\
\cline{1-5}
\multirow{24}{*}{IoT}  &  Smart Things                   & \multirow{22}{*}{\cite{Sivanathan2017}} & Google Home mini          &  \multirow{13}{*}{Our Lab}            \\
                       &  Netatmo Welcome                &                                         & NestCam                   &                                       \\
                       &  TP-Link Day Night camera       &                                         & Amazone Echo              &                                       \\
                       &  Samsung SmartCam               &                                         & Piper Camera              &                                       \\
                       &  Dropcam                        &                                         & Samsung SmartHub          &                                       \\
                       &  Insteon Camera, wifi           &                                         & LifiLab                   &                                       \\
                       &  Insteon Camera, wired          &                                         & iRobot                    &                                       \\
                       &  Withings Smart Monitor         &                                         & TPLink SmartPlug          &                                       \\
                       &  Belkin Wemo switch             &                                         & Provision ISR (camera)    &                                       \\
                       &  TP-Link Smart plug             &                                         & Motorola Hubble (camera)  &                                       \\
                       &  iHome                          &                                         & Foscam                    &                                       \\
                       &  Belkin wemo motion sensor      &                                         & Ring Doorbell             &                                       \\
                       &  NEST Protect smoke alarm       &                                         & Samsung smart home camera &                                       \\
\cline{4-5}
                       &  Netatmo weather station        &                                         & Google Home mini          &\multirow{3}{*}{\cite{lander, impact}} \\
                       &  Withings Smart scale           &                                         & RENPHO Humidifier         &                                       \\
                       &  Blipcare Blood Pressure meter  &                                         & Wyze Cam                  &                                       \\
\cline{4-5}
                       &  Withings Aura smart sensor     &                                         &                           &                                       \\
                       &  LiFX Smart Bulb                &                                         &                           &                                       \\
                       &  Triby Speaker Smart Speaker    &                                         &                           &                                       \\
                       &  PIX-STAR Photo-frame           &                                         &                           &                                       \\
                       &  HP Printer                     &                                         &                           &                                       \\
                       &  Nest Dropcam                   &                                         &                           &                                       \\
\cline{2-3}
                       &  Chromecast Streamer            &   \multirow{2}{*}{Our volunteers}       &                           &                                       \\
                       &  Yeelink Smart Light Bulb       &                                         &                           &                                       \\
\hline

\end{tabular}
\caption{ Seen and Unseen Dataset \label{Tbl:Devices}}
\end{table*}
}
}
}
\section{ Classifier on Traffic Features}
\label{sec:TrafficClassifier}

In this section, we propose a logistic regression classifier that operates on traffic features.   We start by explaining the two-step learning phase (see Section \ref{sec:learning}), which consists of a feature selection followed by constructing an optimized feature set. We then explain the intuition behind the selected features  in Section \ref{sec:intuit}.
In Section \ref{seq:exper_res} we  present the testing phase results that demonstrate its accuracy on the unseen data set. We then discuss some implementation considerations (in Section \ref{sec:implement}).

\ignore{ 

\begin{algorithm}
\caption{The Classification Algorithm}\label{main_classifier}
\begin{algorithmic}[1]
\Procedure{predict}{$\vec{x}$}
\State \text{learned parameters:} $\vec{\theta}, (\vec{\mu}, \vec{\sigma}), \vec{\text{def}}$
\State $x \gets $imputation($\vec{x}, \vec{def}$)
\State $x \gets $normalize($\vec{x}, (\vec{\mu}, \vec{\sigma})$)
\State $prediction \gets \text{sign}( (1, \vec{x}) \cdot \vec{\theta} )$
\State \If {$prediction \leq 0$}
\State {\Return \text{NoT}}
\Else 
\State \Return \text{IoT}
\EndIf
\EndProcedure
\State
\State

\Procedure{imputation}{$\vec{x}, \vec{def}$}
\For {$i \gets 1$ to $||\vec{x}||$}
\If {$x_i$ is undefined}
\State {$x_i \gets $ def$_i$}
\EndIf
\EndFor
\State \Return $\vec{x}$
\EndProcedure
\State
\State

\Procedure{normalize}{$\vec{x}, (\vec{\mu}, \vec{\sigma})$}
\For {$i \gets 1$ to $||\vec{x}||$}
\State {$x_i \gets (x_i - {\mu}_i) / \sigma_i$}
\EndFor
\State \Return $\vec{x}$
\EndProcedure
\end{algorithmic}
\end{algorithm}
}

\begin {table*}[ht]
\begin{center}
\resizebox{\linewidth}{!}
{
\begin{tabular}{|l||l| }
\hline
Layer	& Feature description \\
\hline
\hline

Link-Layer & Number of outgoing packets                                         \\
Link-Layer & Bandwidth (in bytes) of outgoing traffic                           \\
Link-Layer & Average (in bytes) of packets length                               \\
Link-Layer & Average of interleaving time for outgoing packets                  \\
Link-Layer & Standard deviation of interleaving time for outgoing packets       \\
IP & Number of unique interacted endpoints of remote IPs                        \\
IP & Average of the TTL value in outgoing IP packets                            \\
IP & Average of the header length value in outgoing IP packets   \\
IP & Maximum of the header length value in outgoing IP packets   \\
IP & Minimum of the header length value in outgoing IP packets   \\
IP & Count of unique header length values in outgoing IP packets \\
IP & Number of unique outgoing ports                                            \\
IP & Ratio between the number of TCP to UDP  packets                            \\
IP & Number of unique interacted endpoints of remote End-Points (IP $\times$ Ports)    \\
TCP & Maximum TCP window size                                                   \\
TCP & Mean TCP window size                                                      \\
TCP & Minimum TCP window size                                                   \\
TCP & Count of unique TCP window size values                                   \\
TCP & Linear-least-square error for TCP timestamp value                         \\
DNS & Number of unique DNS queries                                              \\
DNS & Number of DNS queries                                                     \\
HTTP & Average length of user-agent field in http requests                      \\

\hline
\end{tabular}}

\vskip 6pt
\caption{List of raw features tested.} \label{tab:raw_features} 
\end{center}
\end {table*}

{\tiny
\begin {table*}[h]
\begin{center}
\resizebox{\linewidth}{!}
{
\begin{tabular}{|l||l|l| }
\hline
Feature name	& Feature description &	F1-score \\
\hline
\hline

window size &	maximum TCP window size &	$0.942$ \\
\# unique DNS reqs  &	number of unique DNS queries &	$0.845$ \\
\# remote IPs &	number of unique interacted  endpoints of remote IPs  &	$0.829$ \\
\# dns reqs &	number of DNS queries &	$0.738$ \\
\# ports &	number of unique outgoing ports & 	$0.658$ \\
bandwidth	& bandwidth (in bytes) of outgoing traffic  & 	$0.601$  \\
pckt count	& number of outgoing packets & 	$0.588$ \\
tcp ts deviation &	 linear-least-square error for TCP timestamp value &	$0.582$ \\
interleaving deviations & standard deviation of interleaving time for outgoing packets  &	$0.576$ \\
tcp/udp ratio & ratio between the number of TCP to UDP  packets &	$0.548$ \\

\hline
\end{tabular}}

\vskip 6pt
\caption{List of features with F1-score above 0.5   (seen  dataset, time slot of 10 minutes). \vspace{-0.6cm}} \label{tab:features} 
\end{center}
\end {table*}
}

\subsection{Learning phase}
\label{sec:learning}
\subsubsection{Feature Selection}
\label{seq:feature_selection}

We tested 22 features from standard protocols (Link-Layer, IP, TCP, DNS, HTTP)  (see Table \ref{tab:raw_features}). We tested every feature that  we thought might be an indicator. We then  \emph{automatically} selected from among them a small set of ten informative  features  that achieved an F1-score above 0.5 for the seen dataset  (see Table \ref{tab:features}). The feature selection was done as follows: the seen data traffic was first divided into sets of  IoT or NoT. For each feature, we analyzed those sets (IoT vs. NoT) using statistical tools (Welch's t-test, ROC curve and AUC calculation \cite{Welch2008, Fawcett2006}) to determine the separation potential for each feature. We narrowed our feature set to the best performing features. Then, we normalized the seen traffic, such that each device had the same number of samples. This helped us deal with the big differences in recording time, where some devices were recorded for weeks and some for hours. In order to represent a wide range of scenarios, we chose representative samples according to bandwidth (low, medium and high).

We calculated the F1-score using 5-fold cross-validation, a common technique in machine learning, to choose features with no over-fitting. Our 5-fold cross-validation method randomly splits the devices in the dataset into 5 independent sets of train (80\%) and test (20\%). Note that no device appears in more than one test set.  Every test was run 5 times, once for each fold, and the results were averaged. We ran this test separately for every classification latency.

For feature information that appears only from time to time in the traffic (e.g., TCP timestamp or user-agent),  we filled slots with  missing values with the  average values of the feature, as learned from the seen dataset.  Therefore, a low F1-score may also indicate that this feature does not appear in most slots.

\begin{figure}[t]
\centering
\psfig{file=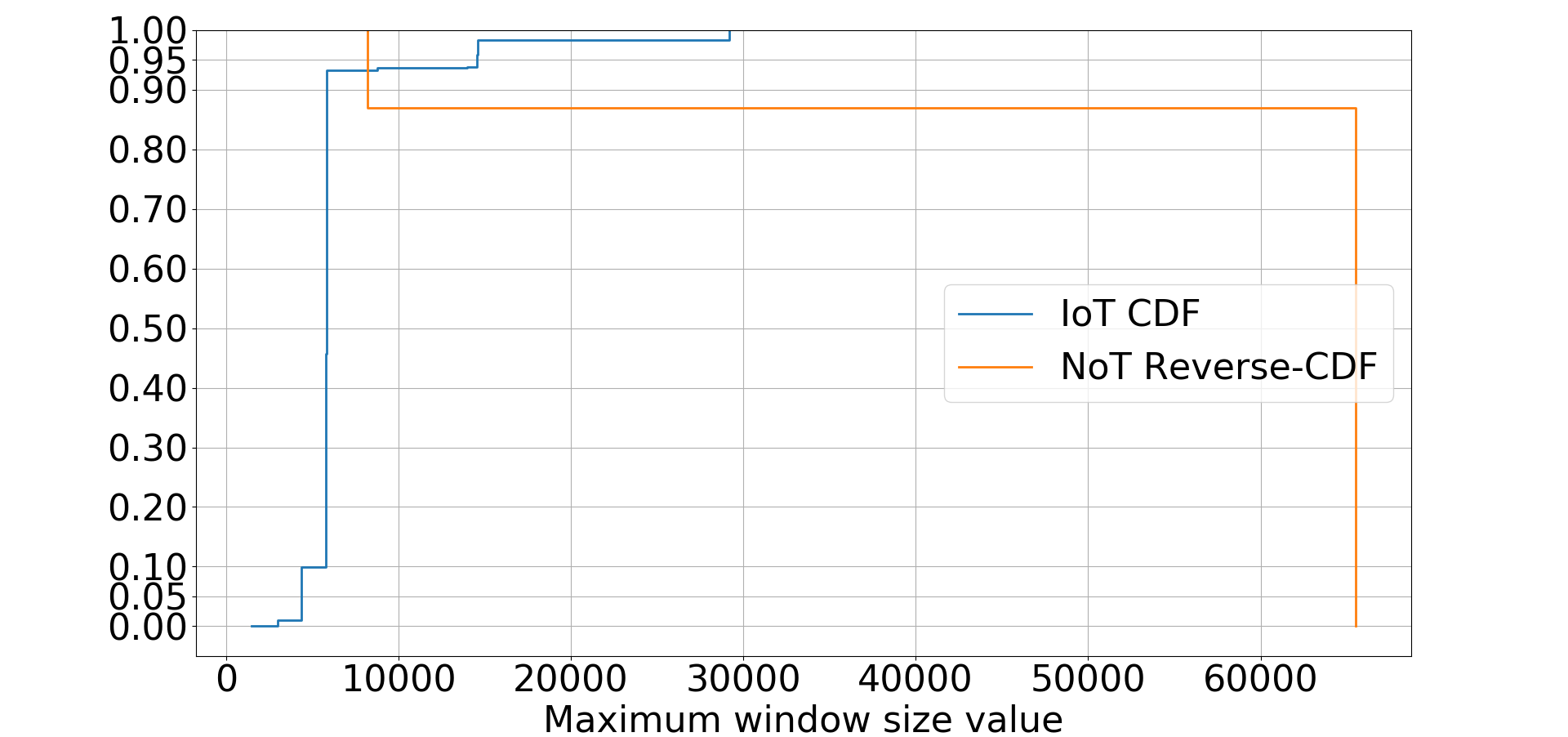,width=3.5in,}
\caption{The values of maximum TCP window size in IoT devices (presented as CDF) and NoT devices (presented as Reverse-CDF), seen dataset, 10-minute time slot. }
\label{fig:winds_cdf}
\end{figure}

\ignore{
\begin{figure}[t]
\centering
\psfig{file=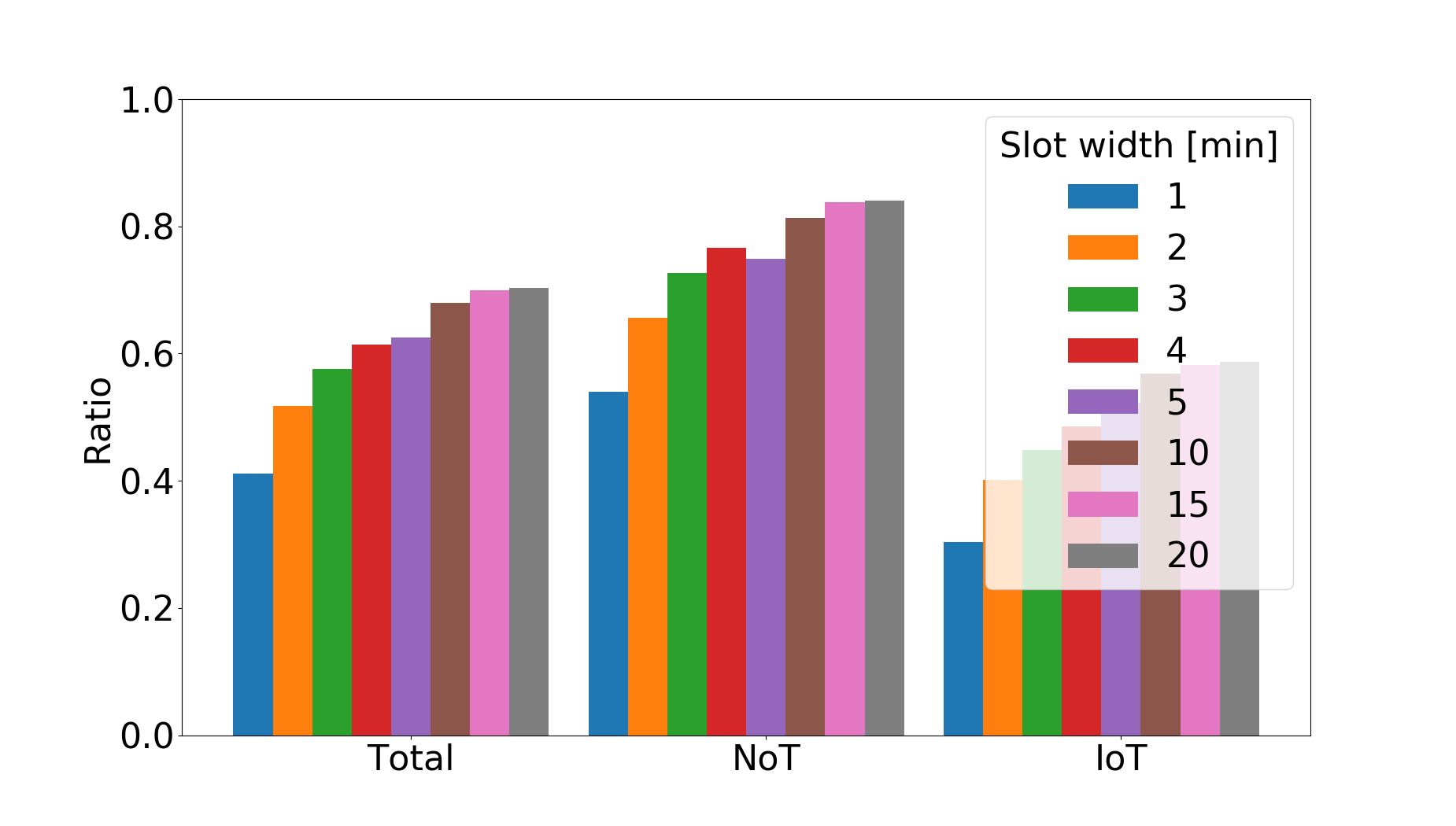,width=3.5in,}
\caption{Percentage of slots that contain TCP-Timestamp field, seen dataset, 10-minute time slot } \label{fig:TCPTS_per}
\end{figure}

\subsubsection{TCP Timestamp Deviation}
The TCP timestamp is a TCP option used to measure RTT. As suggested in RFC 1323 \cite{Jacobson1992}, TCPTS is incremented according to a 'virtual clock' that depends on the device's CPU. This clock has a bias that emerges on TCP packets. Our guess was that less monotonic CPU processing intensity would result in a greater bias for this clock. Thus we assumed that greater bias would be observed for NoT devices than IoT devices. Therefore, we extract the arrival time and TCPTS values from each TCP packet containing this parameter. For each device we calculate linear least square error  on those values. We collect the error as a feature. Figure \ref{fig:TCPTS_per} shows the availability of this optional field in our seen dataset.  
}

\begin{figure}[t]
\centering
\psfig{file=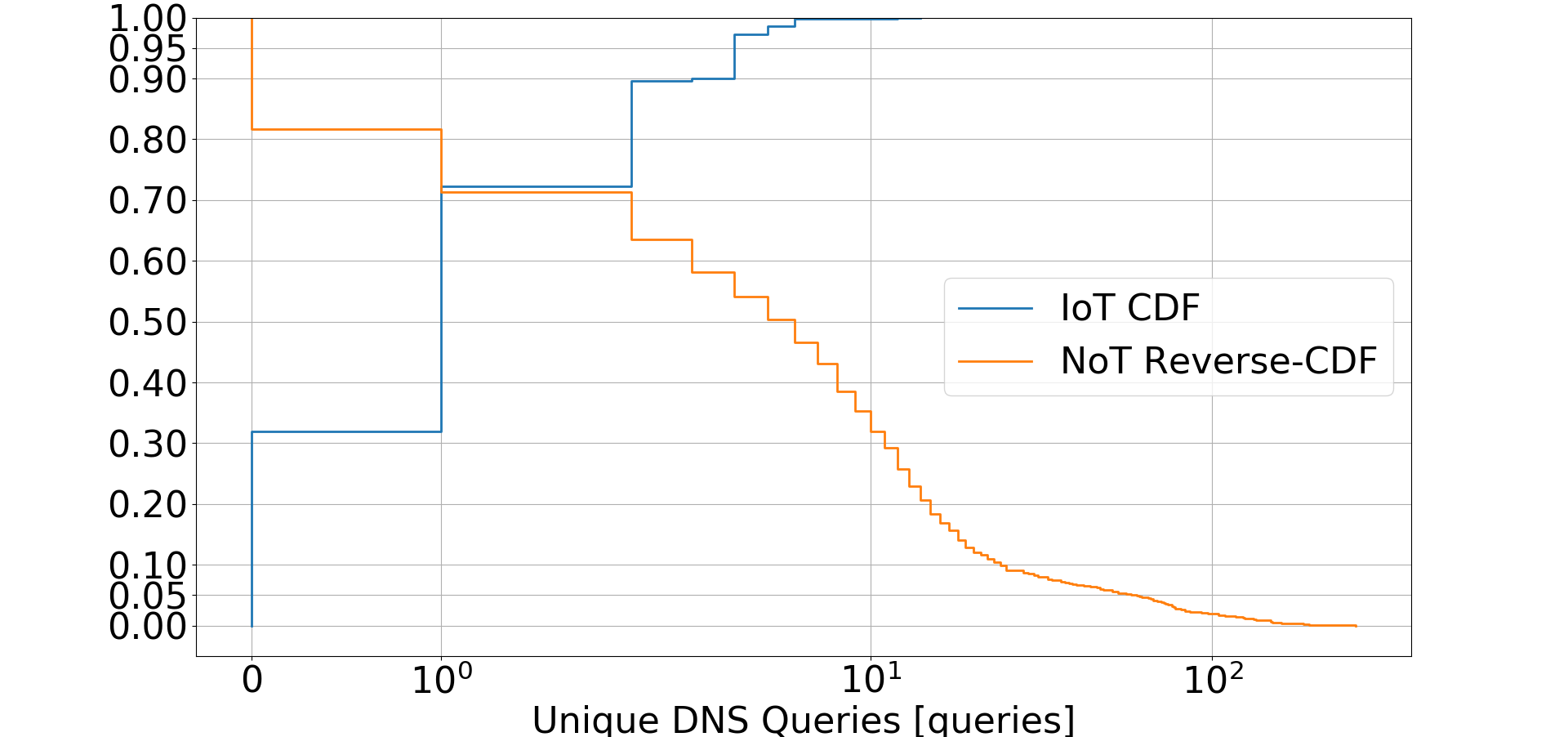,width=3.5in,}
\caption{The number of unique DNS queries in IoT devices (presented as CDF) and NoT devices (presented as Reverse-CDF), seen dataset, 10-minute time slot.} \label{fig:udns_cdf}
\end{figure}

\ignoreInfocom{
\begin{algorithm}
\caption{The Classification Algorithm}\label{main_classifier}
\begin{algorithmic}[1]
\Procedure{predict}{$\vec{x}$}
\State \text{learned parameters:} $\vec{\theta}, (\vec{\mu}, \vec{\sigma}), \vec{\text{def}}$
\State $x \gets $imputation($\vec{x}, \vec{def}$)
\State $x \gets $normalize($\vec{x}, (\vec{\mu}, \vec{\sigma})$)
\State $prediction \gets \text{sign}( (1, \vec{x}) \cdot \vec{\theta} )$
\State \If {$prediction \leq 0$}
\State {\Return \text{NoT}}
\Else 
\State \Return \text{IoT}
\EndIf
\EndProcedure
\State
\State

\Procedure{imputation}{$\vec{x}, \vec{def}$}
\For {$i \gets 1$ to $||\vec{x}||$}
\If {$x_i$ is undefined}
\State {$x_i \gets $ def$_i$}
\EndIf
\EndFor
\State \Return $\vec{x}$
\EndProcedure
\State
\State

\Procedure{normalize}{$\vec{x}, (\vec{\mu}, \vec{\sigma})$}
\For {$i \gets 1$ to $||\vec{x}||$}
\State {$x_i \gets (x_i - {\mu}_i) / \sigma_i$}
\EndFor
\State \Return $\vec{x}$
\EndProcedure
\end{algorithmic}
\end{algorithm}
}
\subsubsection{Constructing an Optimized Feature Set}
\label{seq:optimal_set}

Our classifier uses sklearn's StandardScaler function \cite{standard_scaler} in order to standardize feature values and relies on the logistic regression algorithm \cite{Regression1978} when applying classification. In order to test incoming traffic against the classifier, we again imputed any missing value from an average we learned for each feature. Later on, we used  logistic regression on the standardized values in order to get a classification result.

\ignoreInfocom{
apply Algorithm \ref{main_classifier} on an extracted feature set $x$. $\theta$ are the trained coefficients for the logistic regression algorithm, $\mu$ and $\sigma$ are vectors of the trained mean and scale for the standard scaling for each feature type in the sets, and $def$ is a vector used to impute values for each feature type in case of missing values.  $\mu, \sigma$ and $def$ are general vectors for IoT and NoT.

Performance wise, implementing the classifier on traffic features requires finding the number of distinct elements (of the number of unique dns queries and the number of remote IPs). Note, however, that if the number of distinct elements is small, the naive solution would be to restore the last $x+1$ unique elements seen.
 Since the threshold of the number of unique elements is small, this is a practical solution. One may also apply algorithms that approximate the number of distinct elements, such as \cite{superspreaders:NDSS2005, Bandi:ICDCS2007,Locher:spaa2011}.
}

In constructing a machine learning model, we  chose a combination of features that optimizes prediction rates, for each classification latency separately. We applied our learning using 5-fold cross-validation, as proposed in Section \ref{seq:feature_selection}

To learn the optimal number of features and the optimal features themselves, we used a greedy algorithm. Our goal was not only to optimize the F1-score for each classification latency but also the number of selected features. We used a parameter $\alpha$ (set to $1\%$) as a threshold in order to prefer a smaller vector size over a larger one with a tiny performance gain (less than $\alpha$).  Table \ref{tab:Best_features} shows the optimal combinations and their averaged F1-score over the 5-fold cross-validation for each classification latency (on the seen dataset).

\begin {table*}[h]
\begin{center}
\resizebox{\linewidth}{!}
{
\begin{tabular}{|l||l|l| }
\hline
Classification Latency & F1-Score & Feature set \\
\hline
\hline
1 min & 89.52\% & window size, \# unique DNS reqs, tcp/udp ratio, pckt count \\
5 min & 95.41\% & window size, \# unique DNS reqs, \# remote IPs \\
10 min & 96.33\% & window size, \# unique DNS reqs, \# remote IPs, interleaving deviations \\
20 min & 96.48\% & window size, \# unique DNS reqs, \# remote IPs \\

\hline
\end{tabular}}

\vskip 6pt
\caption{Best feature sets for each classification latency with the F1-Score over cross-validation data (seen dataset)\vspace{-0.5cm}} \label{tab:Best_features} 
\end{center}
\end {table*}

\subsection{Intuition}
\label{sec:intuit}
We then tried to understand the reason  behind the  dominant selected features: window size and number of unique DNS requests. We noticed that IoT device hardware is not as well equipped as NoT devices hardware, and have small buffer size for TCP stack and therefore commonly has a smaller TCP window size  \cite{C.GomezJ.CrowcroftScharf2018}.
  Figure \ref{fig:winds_cdf} compares the CDF of the IoT window size values to the reverse-CDF of the NoT window size values. This comparison shows the separability over this feature. This feature is highly available, visible and unencrypted. All of our devices had TCP traffic in their time slots.
  
In addition, IoT  devices connects to limited endpoints (mostly vendor cloud servers), and thus have fewer unique DNS requests, remote IPs and ports (a similar observation was made in \cite{Sivanathan2017}). 
Figure \ref{fig:udns_cdf} compares the CDF of the number of unique DNS queries of IoT devices to the reverse-CDF for NoT devices.  Note that if there is no DNS traffic, this is also data, and the value of the feature for that slot will be zero. 

 The classifiers that work on time-slots from $5$ minutes and above used  the number of unique remote IPs in addition to the number of the unique DNS requests.
We suspect that these numbers differ since we capture the traffic in some slots, i.e., in the middle of the device operation,  we might not capture the DNS queries that resolve the IPs. Thus the number of unique remote IPs adds information.

\subsection{Testing phase}
\label{seq:exper_res}

After training our models, we validated the classifiers against the \emph{unseen} dataset (see Table \ref{tab:final_results}).  Again, we considered all the time-slots and average the results per device. We received a good F1-score ($91.54\%$) for 1 minute time slot and very high F1-scores (above $96.5\%$) from 5 minutes time slot and above. \ignore{ Surprisingly, this result is even a little better  than  the results on the \emph{seen} dataset.}
\ignore{ As expected, adding information (longer classification latency) to the model usually increases the success rate.}

\begin{figure}[t]
\centering
\psfig{file=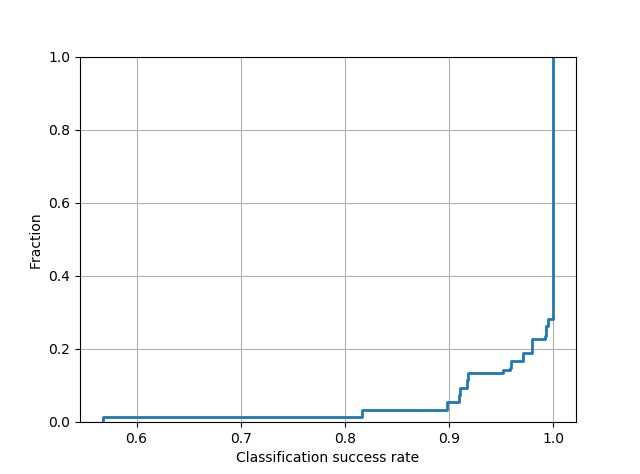,width=3.5in,}
\caption{ CDF of classification success rate of devices,  classifier on traffic features, $10$ minute time-slot,  unseen dataset.  } \label{fig:Traffic_devices}
\end{figure}

In Figure  \ref{fig:Traffic_devices}  we present the CDF of the classification success rate,  defined as the fraction of time-slots, with correct classification of a device. 
For a given class classification success rate $x$, the graph shows the fraction of devices with a successful  classification rate smaller than or  equal to $x$.
Except for one device, all the inconsistently classified  devices were  classified in more than  $82\%$ of the time slots in the same correct way. The most inconsistent devices were  NoT devices, such as Apple iPAD, Samsung Android, Win 10 and Win 7. Devices are incorrectly classified  when the window size is not informative enough, and the NoT device is not very active in that time slot. To improve the results for  these cases, we present in the next section, a classifier that works on the DHCP information.  This result also motivated us to test a classifier with longer classification latency that uses the majority  in a sequence of time slots. This observation is applied in our unified classifier (see Section \ref{sec:unif_res}).  

\subsection{Implementation Considerations}
\label{sec:implement}
Performance wise, implementing the classifier on traffic features requires finding the number of distinct elements efficiently (e.g., of the number of unique DNS queries and the number of remote IPs). Note, however, that if the number of distinct elements, denoted by $x$,  is small, the naive solution would be to store the last $x+1$ unique elements seen.
 In our retrieved classifier the threshold of the number of unique elements was small (less than 15), and hence this is a practical solution. Another possible implementation is to  apply algorithms that approximate the number of distinct elements, as was done in \cite{superspreaders:NDSS2005, Bandi:ICDCS2007,Locher:spaa2011}.

\section{DHCP  based Classifier}
\label{sec:DHCP}

  In this section, we present a decision tree based on DHCP information. For devices that are configured to use the DHCP,  IoT and NoT devices use the protocol to notify  the router of their existence in the network with some information about the device.

\begin{figure}[t]
\centering
\psfig{file=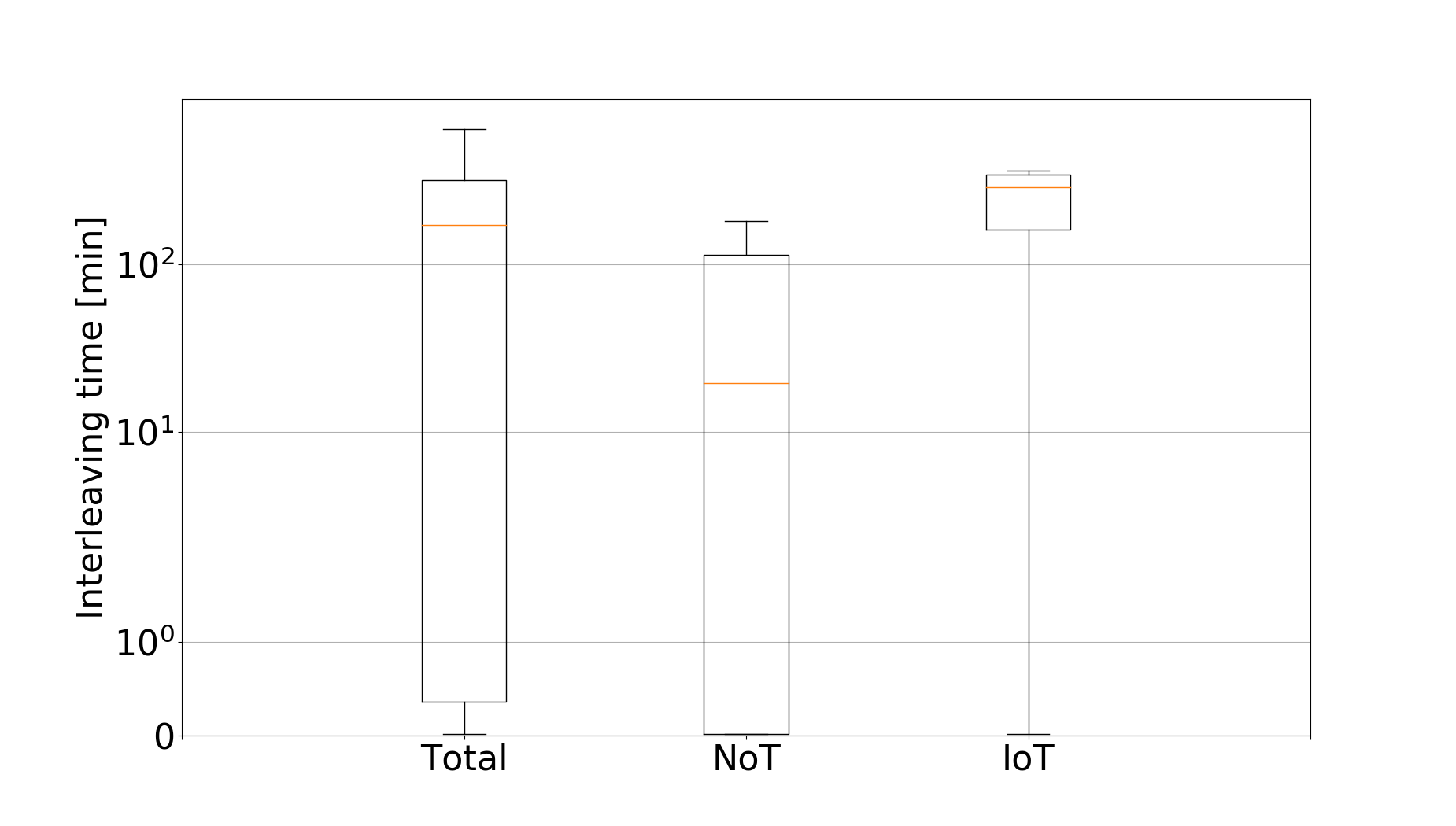,width=3.5in,}
\caption{Interleaving time distribution of DHCP packets on the seen dataset} \label{fig:DHCP_diffs}
\end{figure}

  The  DHCP classifier has some inherent drawbacks:  First, DHCP packets are not available in certain network configurations such as static-IP and in some IPv6 networks due to SLAAC (stateless address auto-configuration \cite{rfc4862}).
Second, classification requires the use of DPI, which is costly in terms of CPU.
Third, DHCP packets are sparsely available, since DHCP traffic occurs when a device connects to the network or renews its IP.   Figure \ref{fig:DHCP_diffs} represents the interleaving time distribution. The median between   consecutive DHCP packets is about 3 hours in our dataset.

Nonetheless, on DHCP enabled networks, this can be overcome and DHCP packets can be retrieved instantly, if active requests can be taken. We can actively disconnect all devices in the network (for example, by running the aireplay-ng tool\cite{aireplay-ng}). This action causes every device to renegotiate and triggers DHCP traffic, and hence we can retrieve the DHCP information in less than three seconds.  With the appropriate credentials, it might be possible to use tr-69 \cite{tr69} (the common  protocol that is used by ISPs to manage and operate the home-routers) to retrieve the required values from the router.~\footnote{Hostname and vci are mandatory in TR-69, other DHCP values are optional in TR69.} 
 
 In the next sections, we explain the learning phase of the decision tree on DHCP information,  the intuition behind the retrieved tree, the testing results on the unseen data, and some implementation considerations.
 
 \subsection{Learning Phase}
 
 In order to construct the classifier automatically, we collected all possible information  from DHCP packets, obtained from five fields: hostname, vendor-class ID (vci), parameter-request-list (prl), maximum-dhcp-size and message-types. 
 We created a list of labels (words/values/numeric values) using the following algorithm: 
For the hostname and vci fields, which contain concatenations of words, we extract labels by splitting those values into labels separated by delimiters (such as ,./\_-+), while filtering numbers. For the parameter-request-list, which is a list of identifiers, we add the identifiers as labels. For maximum-dhcp-size and message-type, which have  numerical values, we add the numerical values as labels. We then construct a binary vector according to those labels for each device. Every i-th bit in a vector represents the fact that the i-th label exists.

 \ignore{
 \begin{itemize}
     \item The {\bf hostname} - is the device's name and it usually contains a value configured by the vendor (but can be changed by the user). Note, that we are seeking for a general classifier, hence we cannot learn all the possible IoT vendors, but we can learn all the major NoT vendors, or major OS vendors, such as Mac, Android and MSFT. 
     \item The  {\bf vendor-class ID} - identifies the device's DHCP client program.  The common values we saw are: 'udhcp' - a lightweight client (common for IoT devices), 'dhcpcd' - a feature-full client (common for NoT devices), common for linux/android based systems, and 'MSFT' - microsoft's DHCP client, shown exclusively for windows-based systems.
 \item The {\bf parameter-request-list} - is a list of parameters the host asks from the DHCP-server to supply upon connection setup. 
 \item {\bf maximum-dhcp-size}  -  sometimes vary between device's types WHAT IS IT??. 
 \item The {\ bf  message-types} - WHAT IS IT ??
 \end{itemize}
 }

\ignore{
\begin{figure}[t]
\centering
\psfig{file=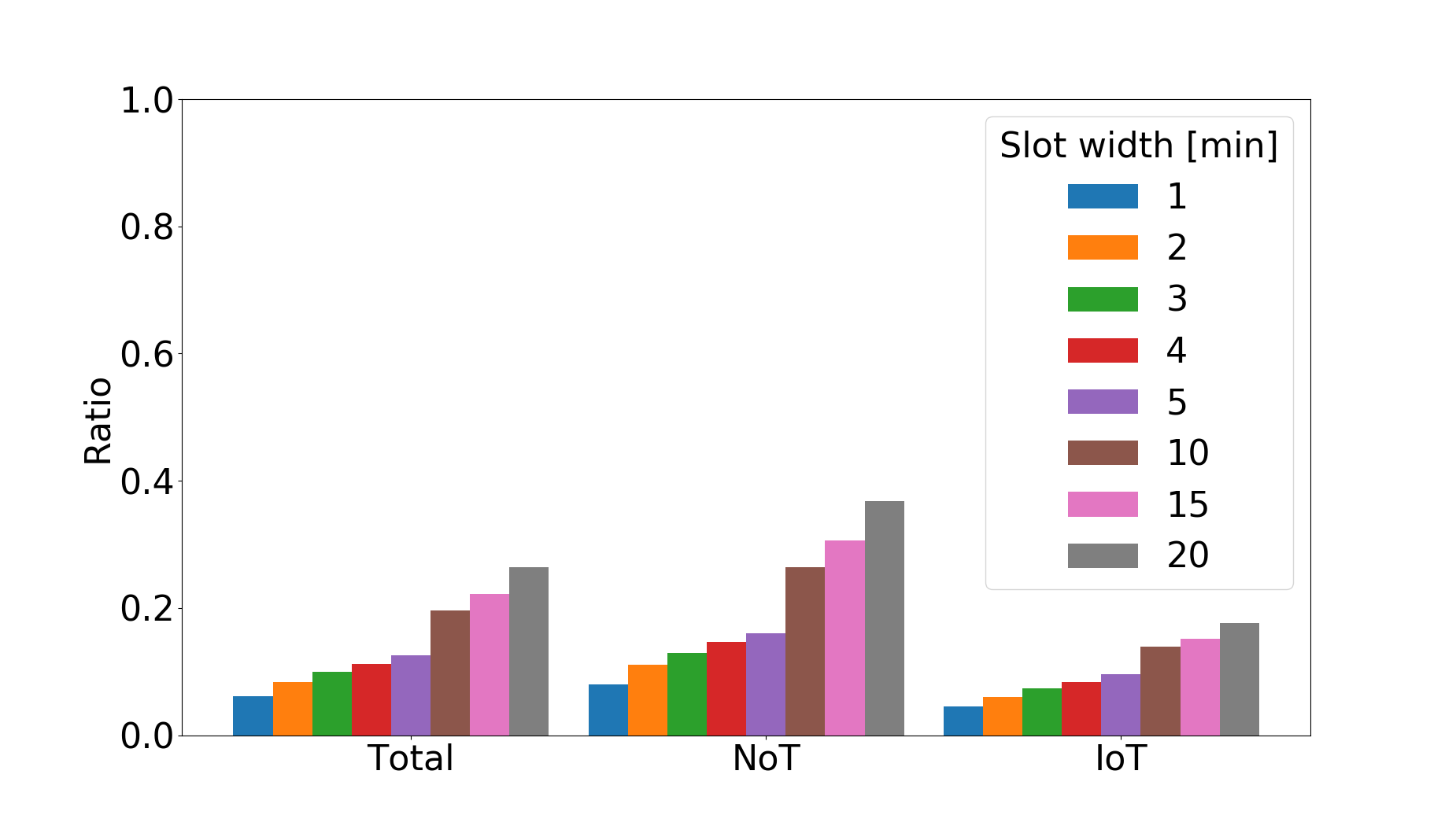,width=3.5in,}
\caption{Percentage of slots that contain DHCP packets on the seen dataset} \label{fig:DHCP}

\end{figure}
}

\begin{figure}[h]
\centering
\psfig{file=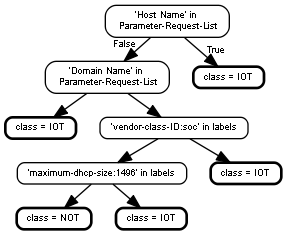,width =3.2in,}
\caption{Decision tree visualization for DHCP -based classifier.\vspace{-0.5cm}} \label{fig:dhcp_tree}
\end{figure}

 \ignore{This binary representation technique is also known as "one hot encoder". }
We trained a decision tree model \cite{PythonDecisionTreePackage} using vectors we constructed according to the seen dataset. We obtained a simple decision tree (see Figure \ref{fig:dhcp_tree}). Fortunately, due to the plurality of devices in the seen dataset, this tree was generic  and did not receive any specific vendor IoT information.

\subsection{Intuition}

The chosen tree uses parameter-request-list (prl) information, which is a list of parameters the device requests from the DHCP server. We saw two dominant values:  the first is the \emph{ hostname} value (tag number 12), which indicates that the device wants its hostname to be assigned by the DHCP server. In our dataset, only IoT devices request it. The second is the \emph{ domain name} value (15), which is used primarily to support easy access to other LAN entities using domain names instead of IP addresses. This value is mainly relevant for NoT devices.

The decision tree also uses vendor-class ID (\emph{vci}) information. The  vci field  mostly contains an information about the type of DHCP client of the device. Some of our IoTs use vci values of SOC (system on chip). We observed a number of other prominent values with high potential, such as 'udhcp' - a lightweight dhcp client (common for IoT), as opposed to 'dhcpcd' - a featureful dhcp client. However, the prl values were more prominent and were chosen by the algorithm.

  We note that the  \emph{hostname} field is the device's name and it usually contains a value configured by the vendor (but can be changed by the user). Note that the decision tree did not choose to use this information, we suspect since there are too many possible IoT vendors and  NoT vendors. 
  
\subsection{Testing Phase}

We tested our model on the unseen dataset and achieved an F1-score of  95.7\%   on devices that use DHCP (see Table \ref{tab:final_results}).   The incorrectly classified  devices  were  Harmon Kardon Invok, RENPHO Humidifier, Ubuntu PC and Homepod of Apple.  
We note that the NoT devices from \cite{cicids2017}  were configured with a static IP, and thus the technique cannot be applied to them.  These devices comprised $10\%$ of our unseen dataset. We classified the devices according to their DHCP packet (one packet is enough), regardless of the classification latency (i.e., without division to time-slots). Thus, the classification latency might be long, if no active request to the router is allowed. We tried to utilize also a random forest algorithm and achieved only slightly better results.  

\ignore{
In order to check the generality of  decision tree  retrieved from the seen  , we decided to learn a  decision tree  from all the dataset, seen and unseen, and compare it to the decision tree that was  learned only from the unseen data. We note that we  received multiple decision trees with the same accuracy, one of them was a decision tree that contains all the nodes in the decision tree  based on the seen data and  with the same  depths. We note that this decision tree that was retrieved from seen all the data, achieved only slightly better accuracy. }

\subsection{Implementation Consideration}
Implementing the DHCP classifier  required  light deep packet inspection, since the required information is in  very specific locations, only in the DHCP packets, so that analyzing one such packet is sufficient. Thus, the DPI would require only very few memory references using known DPI algorithms \cite{Aho:1975:ESM:360825.360855}.

\section{Unified classifier}
\label{sec:unif_res}

As mentioned,  the DHCP information is not always available, but the traffic features classifier is inaccurate for some of the NoT devices in time-slots where the devices are not very active, and when tcp window size information is not inductive enough. 
Hence, we created a unified classifier using the traffic features classifier on different time slots and the DHCP classifier to improve accuracy. 

\begin{figure}[t]
\centering
\psfig{file=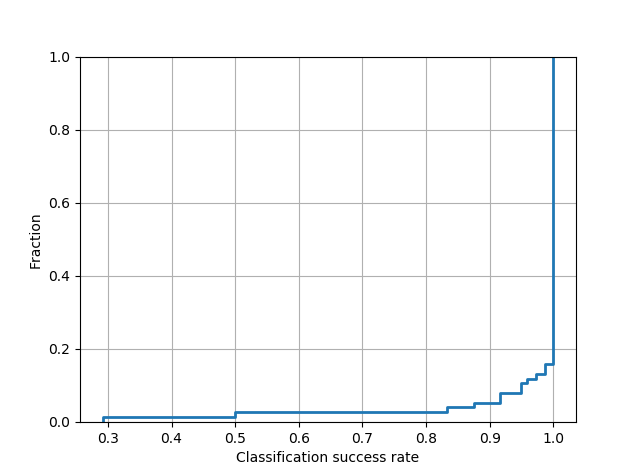,width=3.5in,}
\caption{ CDF of a classification success rate of devices, unified classifier, $20$ minute time-slot, unseen dataset. } \label{fig:DHCP_devices}
\end{figure}

The unified classifier was  heuristically created.  
We focused on a classification time of 20 minutes. For 20 minutes we checked four   5-minute traffic feature classifiers, two 10-minute classifiers, and one  20-minute classifier. Then we combined the classifying results based on traffic features with the DHCP classifier result, and weighted the result as two classifiers if DHCP information exists (as a tie breaker). We classified according to the majority (of nine classifiers). The unified approach slightly improves the accuracy, acheiving an  F1-score of $99.04$;  see Table \ref{tab:final_results} for comparison. We present in  Figure \ref{fig:DHCP_devices} the analysis of the classification success rate per device.

\section{Conclusion \ignore{and Future Work}}
In this paper, we show that it is possible to classify devices as IoT or NoT with short classification latency using simple and efficient classifiers. Understanding whether a device is IoT or NoT is crucial for visibility and security. Our classifiers are able to classify devices that were not seen in the learning phase. This is an important property of our classifier, since there are no datasets that can cover the huge variety of devices, especially IoT devices.  A key benefit of our classifiers is that we can explain the intuition behind the learned classifiers.  The unified classifier code was published in our github repository  \cite{GITHUB}, for use and comparative study by the community. 

A limitation of our research is that we did not focus on borderline IoT devices (such as Android TV and Echo Show). A further study should be performed on the ability to identify this borderline category.

\ignore{
One limitation of our paper is that our classification algorithms resulted from empirical findings on our database, and thus depends on the variety of devices in the dataset. While our database is large in comparison to previous works in the field, we are still cannot cover the huge variety of devices, especially IoT devices. 
Moreover, the IoT field is in an early stage, and still evolving. Thus, with time, the properties of IoT devices might change. However, since we use ML approaches to obtain our algorithms, our ML methods can be used on new datasets and obtain new classification algorithms. We believe that our two proposed classifiers, will still be relevant over time, since they capture the unique properties of IoT devices as compared to general purpose devices. }

 \ignore{
\section{Acknowledgments}
The IoT devices collected from the USC-LANDER IoT-Bootup-Traces-20181107 dataset \cite{lander} were provided by the USC/LANDER project. http://www.isi.edu/ant/lander.
Support for the USC/ LANDER project is provided by the U.S. Department of Homeland Security, Science and
Technology Directorate, IMPACT program.
 }

\bibliographystyle{acm}
\bibliography{MyCollection.bib}

\end{document}